\documentclass[aps,showpacs,preprint,superscriptaddress]{revtex4}
\usepackage{amsmath,amssymb,amsfonts}
\usepackage[english]{babel}
\usepackage{graphicx}
\usepackage{dcolumn}
\usepackage{bm}
\usepackage{color}

\begin{document}

\title{Dynamical constants for  electromagnetic fields with  elliptic-cylindrical
symmetry}

\author{B. M. Rodr\'{\i}guez-Lara}
\email[]{bmlara@fisica.unam.mx}
\affiliation{Instituto de F\'{\i}sica, Universidad Nacional Aut\'{o}noma de M\'{e}xico,
Apdo. Postal 20-364, M\'{e}xico D.F. 01000, M\'{e}xico.}

\author{R. J\'auregui}
\email[]{rocio@fisica.unam.mx}
\affiliation{Instituto de F\'{\i}sica, Universidad Nacional Aut\'{o}noma de M\'{e}xico,
Apdo. Postal 20-364, M\'{e}xico D.F. 01000, M\'{e}xico.}

\date{\today}

\begin{abstract}
Taking into account the characteristics of a free scalar field in
elliptic coordinates, a new dynamical variable is found for the free
electromagnetic field. The conservation law associated to this
variable cannot be obtained by  direct application of standard
Noether theorem since the symmetry generator is of second order.
Consequences on the expected mechanical behavior of an atomic system
interacting with electromagnetic waves exhibiting such a symmetry
are also discussed.
\end{abstract}

\maketitle
\section{Introduction}
Symmetries in physical systems manifest as dynamical constants. In
this way, via Noether theorem, the homogeneity and isotropy of free
space is directly related to the conservation of linear momentum
${\vec P}$ and angular momentum $\vec J$ respectively, and time
homogeneity is related to energy ${\cal E}$ conservation. For a
system of particles interacting through  intermediate fields, global
conservation laws are usually translated, via the equations of
motion, into local conservation laws describing the interchange of
dynamical variables between  particles and  fields. Under proper
circumstances, the effects of macroscopic materials on the fields
are synthesized into boundary conditions for the fields. These
boundary conditions define a set of field configurations (modes)
that satisfy them. The mean value of the different dynamical
variables can be evaluated for each mode inside the region where
boundary conditions are imposed. If this value of the dynamical
variable can change in time just through its flow on the boundaries,
a conservation law for the dynamical variable associated to the
field holds. For instance, the geometry of wave guides define the
electromagnetic (EM) modes and the parameters which characterize
them are linked to dynamical properties of the field. For
rectangular symmetry, the cartesian wave vector $\vec k$ is related
to the momentum $\vec P$ of the EM field, the frequency to the
energy ${\cal E}$ per photon and the polarization to the spin
angular momentum along $\vec k$, $S_z$. For cylindrical symmetry the
energy ${\cal E}$ per photon, linear momentum  $\hat P_z$, orbital
momentum along the symmetry axis $L_z$ and helicity $S_z$, define
the parameters that characterize the electromagnetic Bessel modes:
frequency ${\omega}$, wave vector component along the symmetry axis
$k_z$, azimuthal integer \textsl{m} and polarization $\sigma$. As
shown twenty years ago, Bessel modes \cite{durnin}, as well as other
electromagnetic configurations  coinciding with  modes inside  wave
guides of a given geometry,
 can be generated approximately in free space by interferometric means.
 This has lead to the possibility of creating, in the quantum realm, photons with a
variety of quantum numbers which, in fact, can be entangled
\cite{kwiat}.

The purpose of this work is to study the dynamical properties of
electromagnetic waves in elliptic-cylindrical coordinates. The
corresponding modes are known as Mathieu fields.  The cylindrical
symmetry leads, under ideal conditions, to propagation invariance
along the symmetry axis.  Four of the five  dynamical variables
behind the parameters that characterize Mathieu modes can be
directly identified. The symmetry generator behind the fifth
parameter is trivially obtained from the wave equations but its
relation to the electromagnetic dynamical variable is not, since the
symmetry generator is of second order. We shall give an explicit
expression for this dynamical variable.

The zeroth order Mathieu beams were first
 generated in free space by an annular slit illuminated
with a strip pattern produced by a cylindrical lens \cite{jgv2};
higher order Mathieu beams have already been generated by
holographic means \cite{jgv3}. Given the increasing use of light to
control the motion of atomic systems and microparticles we shall
also make an analysis of the behavior of an atom in a Mathieu field.
Emphasis will be given on the dependence of mechanical effects on
the values of that parameter and the associated dynamical variable.

\section{A massless scalar field in elliptic coordinates.}
Elliptic-cylindrical coordinates are defined by the transformations
\cite{lebedev}
\begin{eqnarray}
x + \imath y &=& f \cosh(u + \imath v), \quad u\in[0,\infty),\quad v\in[0,2\pi) \nonumber \\
z &=& z,
\end{eqnarray}
where the real valued constant $f$ is half the interfocal distance
and the coordinates $u$ and $v$ are the so called
 radial- and angular-like variables, while $z$ is the axial variable.
 Unitary vectors related to a given coordinate $x$ will be written $\hat{e}_{x}$.
The relevant scaling factor for this coordinate system is
\begin{equation}
h = h_{u}= h_{v}= f\sqrt{\left(\cosh 2u  - \cos 2v\right)/2}.
\end{equation}
Considering the solution of the wave equation under the
assumption of propagation invariance along the z-direction,
\begin{equation}
\nabla^2 \Psi = \partial^2_{ct}\Psi, \quad \quad \Psi(\vec{r},t) =
\psi(\vec{r}_{\perp})e^{\imath (k_z z -\omega t)},
\end{equation}
Helmholtz equation is obtained---partial derivatives with respect to a
 given variable $x$ are compactly denoted by $\partial_x$, also the notation
 $\partial_{c t} = \frac{1}{c} \partial_{t}$ will be used.
 In elliptic coordinates, Helmholtz equation takes the form
\begin{eqnarray}
\left\{ \frac{1}{h^{2}} \left( \partial_{u}^{2} + \partial_{v}^{2}
\right) + k_{\perp}^{2} \right\} \psi(u,v) = 0,\quad\quad
k_{\perp}^2= \frac{\omega^2}{c^2} -k_z^2 .\label{eq:helmholtz}
\end{eqnarray}
This equation is separable, $\psi(u,v) =U(u)V(v)$, and yields the
set of differential equations
\begin{eqnarray}
\left\{ \partial_{u}^{2} - b + 2 q \cosh 2u   \right\} U(u)= 0,
 \label{eq:RadialMDE} \\
\left\{ \partial_{v}^{2} + b - 2 q \cos 2v   \right\} V(v)= 0,
\label{eq:AngularMDE}
\end{eqnarray}
known as the modified and ordinary Mathieu equations, in that order,
which will be called radial and angular Mathieu differential
equations from now on. The real constant $q$ is related to both half
the interfocal distance $f$ and the magnitude of the
perpendicular component of the wave vector  $k_{\perp}$ by
\begin{eqnarray}
q = \left(f k_{\perp}/2 \right)^{2}. \label{eq:q}
\end{eqnarray}
From a field theoretical point of view, $q$ is directly related to
the perpendicular momentum carried by the wave $\psi$.  For a given
value of $q$, the possible values of $b$ compatible with the
boundary condition $V(v + \pi) = V(v)$ or $V(v+2\pi) = V(v)$ are
called the characteristic values \cite{abramowitz,mclachlan}. They
are usually ordered in ascending values and renamed  $a_n$ ($b_n$)
for even, $p=e$, (odd, $p=o$) solutions. The parity of the order $n$
determines if the function is $\pi$- or $2 \pi$-periodic for even or
odd order $n$ respectively. The mathematical set
$\{\psi(u,v)^{(p,n,q)}\}$ is complete and orthogonal \cite{jgv}.

From Mathieu radial (\ref{eq:RadialMDE}) and angular (\ref{eq:AngularMDE})
equations, it is straightforward to construct an operator that shares
eigenfunctions with the squared transverse momentum,
$$
 \mathbb{B}~ \psi(u,v)  = b ~ \psi(u,v),$$
 \begin{equation}
 \mathbb{B} = -\frac{f^{2}}{2 h^{2}} \left\{ \cos 2 v \mbox{ }
\partial^{2}_{u} + \cosh 2 u \mbox{ } \partial^{2}_{v} \right\}.
\end{equation}
A physical interpretation of the dimensionless eigenvalue $b$ can be
found writing the
 operator $\mathbb{B}$ in cartesian coordinates
$$\mathbb{B} = -\left(x^2 - \frac{f^{2}}{2} \right)
\partial_{y}^{2} - \left( y^2 + \frac{f^{2}}{2} \right)
\partial_{x}^{2} +
2 x y \partial_{x} \partial_{y} +  x \partial_{x} + y
\partial_{y},$$
\begin{equation}
 \mathbb{B}= \frac{1}{2} \left\{ l_{z+} l_{z-} + l_{z-}
l_{z+} \right\} -\frac{f^2}{2}\nabla_{\perp}^{2} \doteq
l_{[z+}l_{z-]}-\frac{f^2}{2}\nabla_{\perp}^{2} ,\label{eq:B}
\end{equation}
where the operator $l_{z\pm} = - \imath \left[\left( \vec{r} \pm f
\hat{e}_{x} \right) \times \nabla\right]_{z}$, in the context of the
quantum mechanics of a particle, is proportional to the
$z-$component of the angular momentum with respect to either focii
of the elliptic-cylindrical coordinate system.

Previous studies of Helmholtz equation, Eq.~(\ref{eq:helmholtz}),
already report this identification \cite{Traiber1989,leykoo-karen}.
That is the case of the analysis of the spectra of a quantum
elliptic billiard for which $\mathbb{B}$ commutes with the free
particle Hamiltonian \cite{Traiber1989,vanZon1998}. In fact, $
\beta=l_{z+}l_{z-} +2mf^2H$ is a constant of motion of the classical
analogue of this system with $m$ the mass of the particle and $H$
the classical Hamiltonian within the billiard. It has been found
\cite{Traiber1989,berry81} that, if $l_{z+}l_{z-}>0$ then $\beta
>2mf^2E$ and the classical trajectory of the particle repeatedly
touches an ellipse characterized by $\cosh u_{lim} = \beta/2mf^2E$.
While, if $l_{z+}l_{z-}<0$ the trajectory always lies between the
focii touching an hyperbola determined by $\cos v_{lim} =
\beta/2mf^2E$. In that case $\beta$ is a positive number and
$l_{z+}l_{z-}$ has a lower limit given by $-2mf^2E$.

The normalized amplitude of the scalar wave function $\psi(u,v)$ is
illustrated in Fig.~1 for positive and negative values of
$l_{[z+}l_{z-]}$. Notice that, in analogy to the classical particle
problem, for $ l_{[z+}l_{z-]}>0$ ellipses can be observed where the
amplitude is approximately constant while for $ l_{[z+}l_{z-]}<0$
hyperbolic patters of similar amplitude are clearly distinguished.

\begin{figure}[ht!]
\includegraphics[width= 0.9 \textwidth]{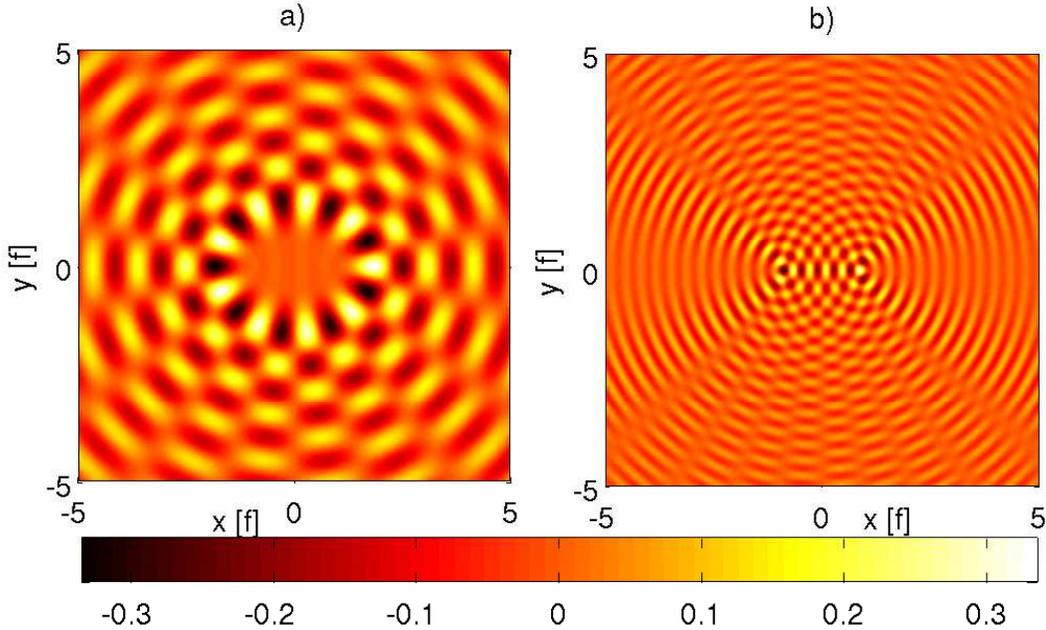}
\caption{\label{fig:MatEven}(Color Online) Normalized even Mathieu's
functions {\bf a)} $\psi(u,v)^{(p=e,n=9,q=10)}$ {\bf b)}
$\psi(u,v)^{(p=e,n=9,q=80)}$. The corresponding values of the
parameter $b$ are $81.6283$ and $124.1067$,  so that $
l_{[z+}l_{z-]}=61.6283$ and $ l_{[z+}l_{z-]}=-44.1067$ respectively.
The half focal distance establishes length units, $f=1$.}
\end{figure}

\section{Dynamical variables of an elliptic electromagnetic mode.}
Given a complete set of scalar solutions $\{\Psi_{\kappa}\}$ of the
wave equation, a joint set of complete electromagnetic modes can be
obtained by considering the scalar functions $\Psi_{\kappa}$ as Hertz
potentials \cite{Stratton41,nisbet57}. In Coulomb gauge, any
solution of the wave equation for the vector electromagnetic field
$\vec A(\vec r,t)$
\begin{equation}
\nabla^2 \vec A(\vec r,t) = \partial^2_{ct} \vec A(\vec r,t)
\end{equation}
can be written as a superposition of  modes
\begin{equation}
\vec{A}_{\kappa}(\vec r,t) = \mathcal{A}^{(TE)}_{\kappa}
\partial_{ct}\vec{\nabla} \times \left( \hat{e}_z \Psi_{\kappa} \right)
+ \mathcal{A}^{(TM)}_{\kappa} \left[ \vec{\nabla}_{\perp} \left(
\vec{\nabla} \cdot \hat{e}_z \Psi_{\kappa} \right) - \hat{e}_z
\nabla_{\perp}^2 \Psi_{\kappa} \right] \label{eq:hertz}
\end{equation}
where  $\kappa$ denotes the labels that characterize a given scalar
solution $\Psi_\kappa$. The constants $\mathcal{A}^{(TE)}_{\kappa}$
and $\mathcal{A}^{(TM)}_{\kappa}$ are proportional to the amplitude
of the transverse electric (TE) and transverse magnetic (TM)
EM fields as can be directly
seen from their connection with the associated electric $\vec{E}$ and
magnetic $\vec{B}$ fields,
\begin{equation}
\vec{E} = - \partial_{ct}\vec{A},\qquad  \vec{B} = \vec{\nabla} \times \vec{A}.
\end{equation}
As usual, although  expressions  may involve complex functions, just
the real part of them define the corresponding physical quantities.
In the case of propagation invariant electromagnetic fields in
elliptic coordinates
\begin{equation}
\vec A_{\kappa}(\vec{r},t) = \mathcal
{A}^{(TE)} \vec{\mathbb{M}} \Psi_{\kappa} + \mathcal {A}^{(TM)} \vec{\mathbb{N}}
\Psi_{\kappa} \label{eq:elliptic}
\end{equation}
where the vector operators are given by the expressions
\begin{equation}
\vec{\mathbb{M}} = \frac{1}{h} \partial_{ct} \left( \hat{e}_{u} \partial_{v} -
\hat{e}_{v} \partial_{u}\right), \qquad  \vec{\mathbb{N}} = \frac{1}{h}
\partial_{z} \left( \hat{e}_{u} \partial_{u} + \hat{e}_{v} \partial_{v}\right) -
\hat{e}_{z} \nabla_{\perp}^{2}.
\end{equation}
As a consequence,
\begin{eqnarray}
\vec{E}_{\kappa}(\vec{r},t) &=& - \mathcal{A}^{(TE)}\partial_{ct}\vec{\mathbb{M}}
 \Psi_{\kappa} - \mathcal{A}^{(TM)}\partial_{ct}\vec{\mathbb{N}} \Psi_{\kappa}, \nonumber\\
\vec{B}_{\kappa}(\vec r,t) &=& \mathcal{A}^{(TE)}\partial_{ct}\vec{\mathbb{N}}
\Psi_{\kappa} - \mathcal{A}^{(TM)}\partial_{ct}\vec{\mathbb{M}} \Psi_{\kappa}.
\end{eqnarray}
In general, it is expected that the electromagnetic modes given by
Eq.~(\ref{eq:hertz}) inherit symmetries of the scalar field with
analogous dynamical variables. For the elliptic-cylindrical case,
invariance under space reflection with respect to the $Y$ axis leads
to parity conservation. Meanwhile, invariance under spatial
translation along the main direction of propagation of the mode is
reflected in the fact that the field momentum-like integral
\begin{equation}
P^{(i,\kappa,\kappa^\prime)}_{z} = \frac{1}{4\pi c}\int d^3x
\left[(\vec E^{(i)}_{\kappa}\times \vec
B^{(i)}_{\kappa^{\prime}})_{z}\right]\qquad i= TE,TM
\label{eq:momentum}
\end{equation}
integrated over the whole space is independent of time. Time
homogeneity  implies that the  energy-like integral
\begin{equation}
\mathcal{E}^{(i,\kappa,\kappa^{\prime})}= \frac{1}{4\pi}\int d^3x
\left[{\vec{E}}^{(i)}_{\kappa}(\vec{r},t) \cdot
\vec{E}^{(i)}_{\kappa^{\prime}} (\vec{r},t)
+\vec{B}_{\kappa}^{(i)}(\vec{r},t) \cdot
\vec{B}_{\kappa^{\prime}}^{(i)}(\vec{r},t)\right]\label{eq:energy}
\end{equation}
is also constant. In fact, $ P^{(i,\kappa,\kappa^\prime)}_{z}$ and
$\mathcal {E}^{(i,\kappa,\kappa^\prime)}$ are proportional with
$ck_z/\omega$ the constant of proportionality. By construction
$\hbar k_{\perp}=\hbar\sqrt{\omega^2/c^2-k_z^2}$  would yield the
magnitude of the transverse component of the momentum which
determine the separation constant $q$, Eq.~(\ref{eq:q}).

Standard quantization rules require a proper normalization selection
for the EM modes so that each photon carries an energy
$\hbar\omega$. Using Eq.~(12a) from Ref.~ \cite{inayat}
\begin{equation}
\int_{-\infty}^\infty dz\int_0^\infty du\int_0^{2\pi} dv
h_uh_v\Psi_\kappa(u,v,z)\Psi_\kappa^\prime(u,v,z) =
2\pi^2f^2s_\kappa\delta(k_z-k_z^\prime)\delta(q-q^\prime)\delta_{n,n^\prime}
\end{equation}
\begin{equation}
\begin{array}{lcl  lcl}
s_{e,2n,q}&=&\frac{V_{e,2n,q}(0)V_{e,2n,q}(\pi/2)}{A_0^{(2n)}},&
s_{e,2n+1,q}&=&-\frac{V_{e,2n+1,q}(0)V^\prime_{e,2n+1,q}(\pi/2)}{q^{1/2}A_1^{(2n+1)}}, \\
s_{o,2n+2,q}&=&\frac{V_{o,2n+2,q}^\prime(0)V^\prime_{o,2n+2,q}(\pi/2)}{qB_2^{(2n+2)}},&
s_{o,2n+1,q}&=&\frac{V^\prime_{o,2n+1,q}(0)V_{o,2n+1,q}(\pi/2)}{q^{1/2}B_1^{(2n+1)}},
\end{array}
\end{equation}
where $A^{(n)}_{m}$ and $B^{(n)}_{m}$ are the standard Mathieu
coefficients \cite{abramowitz,mclachlan}, this normalization is
trivially performed. Defining now the generalized {\it number
operator}:
\begin{equation}
\hat {N}_\kappa^{(i)} = \frac{1}{2} \Big(\hat a^{(i)\dagger}_\kappa
\hat a^{(i)}_\kappa + a^{(i)}_\kappa \hat a^{(i)\dagger}_\kappa
\Big),\qquad [a^{(i)}_\kappa,a^{(j)\dagger}_{\kappa^\prime}] =
\delta_{i,j}\delta_{\kappa,\kappa^\prime}
\end{equation}
the quantum energy and the momentum  along $z$ operators take the
form:
\begin{equation}
\hat {\cal E}=  \sum_{i,\kappa}\hbar\omega ~\hat{N}_\kappa^{(i)} ,
\qquad \hat {\cal P}_z=  \sum_{i,\kappa}\hbar k_z
~\hat{N}_\kappa^{(i)},
\end{equation}
allowing the identification of $\hbar k_{z}$ and $\hbar \omega$ with
the photon momentum along $z$ and the photon energy, in that order.

  For scalar fields and in the case
of spacetime continuous symmetries, the generators of infinitesimal
transformations become good realizations of the corresponding
dynamical operator. In that sense, the rotation like operator $
\mathbb{ B}$ can be related to the product of angular momenta
$l_{[z+}l_{z-]}$ and the eigenvalue equation
$\mathbb{B}~\Psi_{\omega,k_z,p,n}
 =b~\Psi_{\omega,k_z,p,n}$  is interpreted as  a manifestation
of the scalar wave function $\Psi_{\omega,k_z,p,n}$ carrying a well
defined value of that angular momentum product.

  The standard procedure to find the electromagnetic analogue
of $l_{[z+}l_{z-]}$ would be to apply Noether theorem to the field
Lagrangian
\begin{equation}
\mathcal{L}_{EM} = (1/4)(\partial_{\mu} A_{\nu} - \partial_{\nu}
A_{\mu})(\partial^{\mu} A^{\nu} - \partial^{\nu} A^{\mu})\label{eq:lagrangian}
 \end{equation}
 for the  transformation generated by $\mathbb{B}$ on space variables
 and on the electromagnetic fields $A_\mu$. Standard Noether theorem
 \cite{bogoliubov} concerns first order
  differential operators as generators of continuous symmetries, so
  that, if under an infinitesimal transformation that modify the coordinates and
  field functions by
  \begin{equation}
 \delta x_\mu = \sum_\nu X_\mu ^\nu \delta \omega_\nu, \quad \quad
 \delta A_\mu = \sum_\nu \varphi_\mu^\nu \delta\omega_\nu,
  \end{equation}
the Lagrangian is left invariant, the current
\begin{equation}
\Theta_\rho^\nu=-\sum_\lambda\frac{\partial {\mathcal L}}{\partial
A_{\lambda,\nu}}(\varphi_{\lambda\rho}-
\sum_{\sigma}A_{\lambda,\sigma}X_\rho^\sigma) - {\mathcal L}
X^\nu_\rho
\end{equation}
has zero  divergence $\partial_\nu \Theta_\rho^\nu=0$. As a
consequence $\Theta_\rho^0$ can be considered as the density of a
dynamical variable whose integrated value over a volume can change
only due to the flux  of the current $\Theta_\rho^i$ through a
surface. For the circular cylindrical problem, the assumption of
isotropy of space  via the effects of the infinitesimal rotation
generator $ [\vec{r} \times {\vec \nabla}]_z$ on $A_\mu$ and $x_\mu$
leads to the identification of
\begin{equation}
 { J}_z =     \frac{1}{4\pi c} \int_{\cal V}   ~
 \sum_i E_i[{\vec r}  \times\vec \nabla]_z  A_i  ~d^3x
+\frac{1}{4\pi c} \int_{\cal V} ~ ({\vec E}\times  {\vec A})_z ~d^3x
,\label{eq:separation}
\end{equation}
as the z-component of the angular momentum of the electromagnetic
field in a volume ${\cal V}$ \cite{bogoliubov}. The first integral
involves the antihermitian differential operator $ [{\vec r} \times
{\vec \nabla}]_z= \partial_{\varphi}$, and is associated to the
orbital angular momentum (notice that the standard
 hermitian angular momentum operator is $\hat L_z = -\imath \hbar
 \partial_{\varphi}$).
 The second integral is independent of the
choice of origin, arises from the field variation $\delta A_\mu$,
and is directly related to the polarization of the field.  It has
been identified with the field helicity \cite{nienhuis,jauregui}. It
can be shown that in the Coulomb gauge \cite{mandel}
\begin{equation}
{\vec J} = \frac{1}{4\pi c}\int_{\cal V} \vec r \times (\vec E \times \vec B)d^3x\nonumber \\
-    \frac{1}{4\pi c} \oint_{\cal S} {\vec E}\big[({\vec r} \times
{\vec A}\big]\cdot d{\bf s}
\end{equation}
where ${\cal S}$ is the surface enclosing the volume ${\cal V}$.

Since the electromagnetic field $A_\mu$ has a well defined
transformation rule under rotations which is independent of the
origin of space coordinates , the second term in
Eq.~(\ref{eq:separation}),
\begin{equation}
{ S}_z=\frac{1}{4\pi c}\int_{\cal V} ~ ({\vec E}\times  {\vec A})_z
~d^3x,
\end{equation}
is the fourth dynamical variable associated to the Mathieu
electromagnetic field. This can be directly verified by substituting
the general expression for the vectors $\vec E$ and $\vec A$ in
terms of the elliptic modes. As usual, for a given value of mode
indices $\kappa$, the helicity $S_z$ is different from zero only if
the amplitudes ${\mathcal A}^{(TM)}$ and ${\mathcal A}^{(TE)}$ are
complex. The concept of circular and linear polarization of Mathieu
waves is analyzed in Ref.~\cite{karen} classically. The quantum
analysis can be carried out in complete analogy with the study in
Ref.~\cite{jauregui} for Bessel fields so that
\begin{equation}
\hat{ S}_z=     \hbar \sum_\kappa \frac{\imath k_z c}{2\omega}
\big(\hat a^{(TE)\dagger}_m  \hat a^{(TM)}_m - \hat a^{(TE)}_m \hat
a^{(TM)\dagger}_m \big).
\end{equation}

 In the problem treated here, the
generator of the transformation $\mathbb{B}$
  is an hermitian second order differential operator
  obviously related to the isotropy of space.
It depends on the position of the two focii of the elliptic
transversal coordinates and in that sense should be analogous to
$L_z$. The proposal is to identify the electromagnetic dynamical
variable related to $\mathbb{B}$ with the integral
\begin{equation}
\mathcal{B}= \frac{1}{4\pi c} \int_{\cal V} ~ \sum_iE_i ~\imath
\mathbb{B}~ A_i ~d^3x , \label{eq:BEM}
\end{equation}
in complete analogy with Eq.(\ref{eq:separation}).

 In order to corroborate that $\mathcal{B}$ is the fifth dynamical
 variable directly associated to Mathieu electromagnetic waves, notice that
\begin{eqnarray}
\sum_iA_i^{\kappa^\prime} \mathbb{B}  A_i^\kappa &=&
 b ~ \vec{A}^{\kappa^\prime} \cdot \vec A^{\kappa}  \nonumber \\
&+& \left[ k^{\prime} k \mathcal
{A}^{(TE)}_{\kappa^{\prime}} \mathcal{A}^{(TE)}_{\kappa} +
k_z^\prime k_{z}\mathcal{A}^{(TM)}_{\kappa^{\prime}}
\mathcal{A}^{(TM)}_{\kappa} \right]  \cdot \left[\vec \nabla \cdot
\vec C - \Psi_{\kappa^\prime}\nabla^2_\bot\Psi_\kappa\right]
\label{eq:barb}
\end{eqnarray}
where the vector $\vec{C} = - 2
\left(\mathbb{M}\Psi_{\kappa^{\prime}} \right) \left( \vec{r} \times
\vec{\mathbb{N}}_{\perp} \Psi_{\kappa}\right) +
\Psi_{\kappa^{\prime}} \vec{\mathbb{N}}_{\perp}
\Psi_{\kappa}$---with $\vec{\mathbb{N}}_{\perp}=\frac{1}{h}
\partial_{z} \left( \hat{e}_{u} \partial_{u} + \hat{e}_{v}
\partial_{v}\right)$.

Evaluating the integral in Eq.~(\ref{eq:BEM}) over a volume
$\mathcal{V}$, using Eq.~(\ref{eq:barb}), the first resulting
term is proportional to $b \vec{E}^{\kappa^\prime}\cdot
\vec{A}^{\kappa}$ involving the same integrals appearing in the
energy-like density, Eq.~(\ref{eq:energy}). The second term defines
a flux of $\mathcal{B}$ through the surface around the integration
volume $\mathcal{V}$. The third term adds up to the last term in the
expression of operator $\mathbb{B}$, Eq.~(\ref{eq:B}).

Eq.~(\ref{eq:barb}) supports the identification of $ \mathcal{B}$ as
the electromagnetic dynamical variable linked to the generator
$\mathbb{B}$.

In the quantum realm the corresponding operator is written
\begin{equation}
\hat{\mathcal{B}}=  \sum_{\kappa}\hbar \left( b + 1 \right)
\hat{N}_\kappa^{(TE)} + \hbar \left( b +
\frac{k_{z}^{2}c^2}{\omega^2} \right) \hat{N}_\kappa^{(TM)} .
\end{equation}
In the paraxial limit, the helicity-like factor $k_{z}c/\omega\sim1$
so that $\hat{\mathcal{B}} \approx \sum_{i,\kappa}\hbar ( b + 1)
\hat{N}_\kappa^{(i)}$.

 Notice that the dynamical variable $\hat{\mathcal{B}}$ for a
photon has units of $\hbar$, although for material particles the
quantum variable associated to $l_{z+}l_{z-}$ has as natural unit
$\hbar^{2}$. The interpretation of a dynamical variable for the EM
field is usually linked to the interchange of this mechanical
variable with charged particles or atomic systems. The measurement
of $\hat{\mathcal{B}}$ is expected to be related to changes in
values of $l_{z+}l_{z-}$ although they have different units. It is
thus essential to clarify how the absorption and/or emission of a
Mathieu photon by a particle alters its motion.

\section{Mechanical effects on atoms.}
Theoretical and experimental analysis on the interaction between
light and microscopic particles, as well as analysis on the
interaction of light and cold atoms, have yield very important
results in the last three decades. In these areas, the use of
structured light beams with peculiar dynamical properties plays an
increasingly important role \cite{review1}. In the particular case of Mathieu-like
beams, it is possible to generate an elliptical orbital motion of
trapped microscopic particles \cite{OAM}. A detailed theoretical
description of this phenomenon requires the exhibition of a clear
link between the observed motion and the parameters that
characterize the beam, which are directly related to the mechanical
properties of the field described here.

The mechanical effects of a Mathieu electromagnetic wave on a cold
atom shall briefly be described under the assumption that the atom
kinetic energy is low enough to be sensitive to light forces but
large enough to admit a description in terms of Newton equations.
The standard semiclassical approach is taken, as in the pioneering
works by Letokhov \cite{leto} and Gordon and Ashkin
\cite{gordon-ash}. In this approximation, a monochromatic
electromagnetic wave describable by a coherent state couples to the
dipole moment of an atom. This dipole moment $\vec \mu_{12}$ is
related to the electromagnetic transitions between the atom levels
that, for simplicity sake, will be taken to have just two accessible
options. The gradient of the coupling $g= \imath \vec \mu_{12}\cdot
\vec E/\hbar=\vert g\vert e^{\imath \phi}$ determines the explicit
expression for the average semiclassical velocity dependent force
through the vectors $\vec\alpha = \vec\nabla\log( \vert g\vert)$ and
$\vec \beta = \vec\nabla \phi$ \cite{gordon-ash}. The nonlinear
Newton equations for Mathieu waves have a rich structure that
deserves a deep study on its own. Here, just some results that
illustrate the relevance of the parameter $b$ are reported.
 Atomic transitions with changes in the atomic internal angular
 momentum $\delta \mathfrak{m} = \pm 1$ are proportional to
 $\hat e_\pm\cdot\vec E (\vec r,t)$. For Mathieu waves, this factor
 is proportional to $ \left( \omega/ c \right) {\mathcal A}^{TE} \pm \imath k_z{\mathcal A}^{TM}$
 reinforcing the interpretation of the latter expression as a signature of circular
 polarization. In order to induce  transitions with $\delta \mathfrak{m}= \pm 1$ with
 equal probability, and to avoid any complication arising from a component
 of the electric field along $z$, it will be assumed that
  ${\mathcal A}^{TM}=0$. Under these conditions,
  \begin{eqnarray}
  \vec\alpha&=& \frac{\hat e_u}{h}\Big[\frac{\partial_v\Psi\partial^2_{uv}\Psi +
   \partial_u\Psi\partial^2_{u}\Psi}{(\partial_u\Psi)^2 +
   (\partial_v\Psi)^2} - \frac{\sinh 2u}{\cosh2u -\cos
   2v}\Big]\nonumber\\
&+& \frac{\hat e_v}{h}\Big[\frac{\partial_u\Psi\partial^2_{uv}\Psi +
   \partial_v\Psi\partial^2_{v}\Psi}{(\partial_u\Psi)^2 +
   (\partial_v\Psi)^2} - \frac{\sin 2v}{\cosh2u -\cos
   2v}\Big],
  \end{eqnarray}
 \begin{eqnarray}
  \vec\beta&=& \frac{\hat e_u}{h}\Big[\frac{\partial_u\Psi\partial^2_{uv}\Psi
  -
   \partial_v\Psi\partial^2_{u}\Psi}{(\partial_u\Psi)^2 +
   (\partial_v\Psi)^2} - \frac{\sin 2v}{\cosh2u -\cos
   2v}\Big]\nonumber\\
&-& \frac{\hat e_v}{h}\Big[\frac{\partial_v\Psi\partial^2_{uv}\Psi -
   \partial_u\Psi\partial^2_{v}\Psi}{(\partial_u\Psi)^2 +
   (\partial_v\Psi)^2} - \frac{\sinh 2u}{\cosh2u -\cos
   2v}\Big].
  \end{eqnarray}
 The expression for the average semiclassical
velocity dependent force \cite{gordon-ash} valid for both
propagating and standing beams is:
\begin{equation}
\langle \vec f \rangle = \frac{\hbar\Gamma\Big[[(\vec
v\cdot\vec\alpha)(1-p)(1+p)^{-1} +\Gamma/2] \vec  \beta + [(\vec
v\cdot \vec\beta) -
\delta\omega]\vec\alpha\Big]}{(1-p^\prime)p^{\prime -1}\Gamma +2\vec
v\cdot\vec\alpha[1-p/p^\prime -p](1+p)^{-1}} \label{eq:force}
\end{equation}
with $\Gamma$ the Einstein coefficient, $\Gamma =
4k^3\vert\vec\mu_{12}\vert^2/3\hbar$, $\delta\omega$ the detuning
between the wave frequency $\omega$ and the transition frequency
$\omega_0$,  $\delta\omega = \omega -\omega_0$, $p =2\vert\vec
g\vert/((\Gamma/2)^2+\delta\omega^2)$ a parameter linked to the
difference $D$ between the populations of the  atom  two levels,
 $D=1/(1+p)$, and finally
$p^\prime = 2\vert g\vert^2/\vert \gamma^\prime\vert^2$, with
$\gamma^\prime =(\vec v\cdot\vec\alpha)(1-p)(1+p)^{-1} =\Gamma/2 +
\imath (-\delta\omega +(\vec v\cdot\vec\beta))$.

This brief study is focused on the  red detuned far off resonance
light case  so that nonconservative terms arising from the velocity
dependence of the force are not dominant \cite{fort}. This regime is
particularly relevant in the context of optical lattices
\cite{optlatt}. Following Ref.~\cite{fort}, the laser beam is
considered with a 67 nm detuning to the red of the 5 $^2S_{1/2}$ - 5
$^2P_{1/2}$ transition at 795 nm; a 6$\cdot10^5$ W/cm$^2$
irradiance is assumed. The trajectories of the atoms are described taking  the
laser wavelength as unit of length and as unit of time the inverse
of the Einstein coefficient $\Gamma$ which is 3.7 $\times
10^{7}s^{-1}$ for the state 5$^2P_{1/2}$ of $^{85}$Rb.

Since Newton equations are highly nonlinear, it is expected that the
motion of the particles will not have a simple structure.
Nevertheless, there are some general characteristics of the atom
motion that can be predicted without making a numerical analysis.
Thus, since the z-dependence of the beam is just on its phase,
$k_zz$, no longitudinal confinement is expected. In the standard
paraxial regime, $k_{\perp} << k_z$, a correspondingly small
transfer of radial momentum to the atom will occur. Under these
conditions, the atom would be confined by transversal light
potential wells whenever its height is larger than the initial atom
kinetic energy. Conversely, if $k_\bot\sim k_z$ the atom may dwell
around different transverse wells. Typical trajectories are shown in
Figs.~2-3 \footnote{ A computer movie of the time evolution can be
found in www.fisica.unam.mx/research/movies }.
\begin{figure}[ht!]
\includegraphics[width= 0.9 \textwidth]{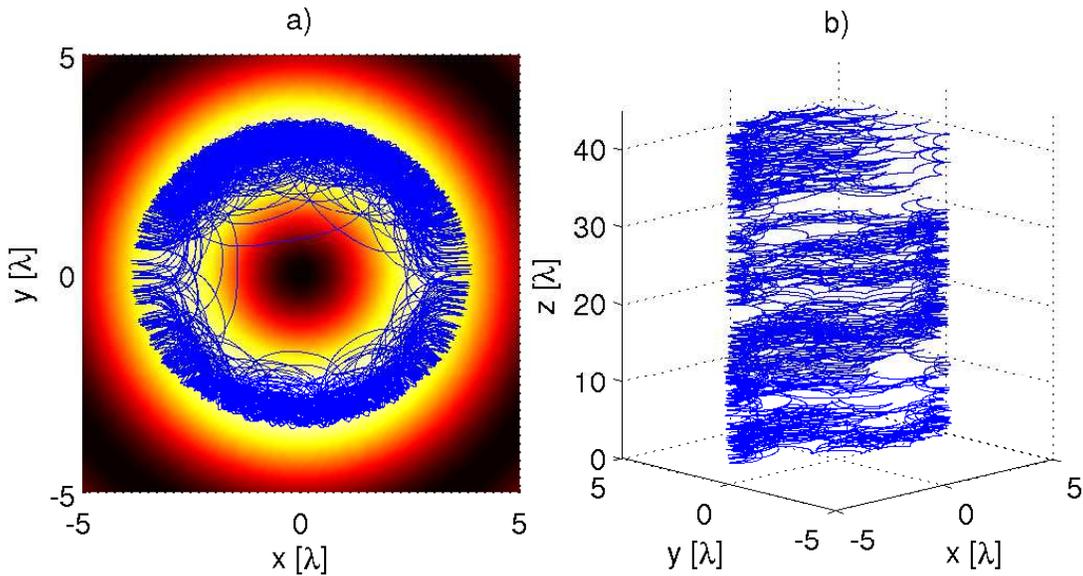}
\caption{(Color online) Trajectory of an atom driven by an even
Mathieu beam of order $n=0$ in the paraxial regime with half focal
distance $f=\lambda$, and beam parameters $q =0.0984$ and $b =-0.0048$.
The initial conditions for the atom are $u_0=1.5 \lambda$,
$v_0=\pi/4$, $z_0=0 \lambda $, $\dot u_0=0.1 \lambda \Gamma$, $\dot
v_0=-0.001 \Gamma$, $\dot z_0 =0.001\lambda\Gamma$.}
\end{figure}

\begin{figure}[ht!]
\includegraphics[width= 0.9 \textwidth]{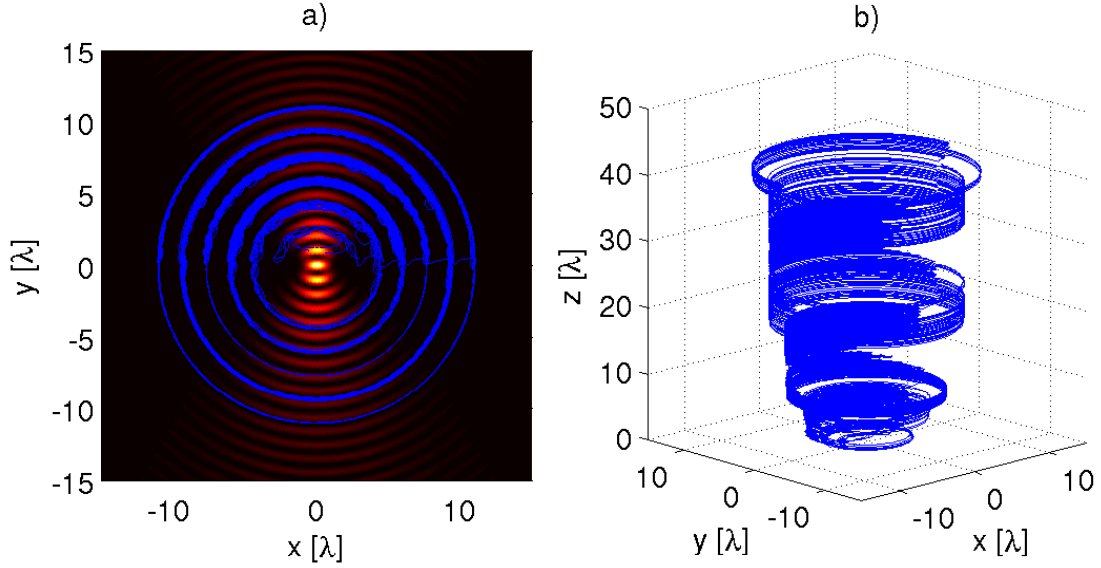}
\caption{(Color online) Trajectory of an atom driven by an odd
Mathieu beam or order $n=1$ with half focal distance $f=\lambda$, and beam
parameters $q = 3.5531$ and $b = -3.5924$. The initial conditions
for the atom are $u_0=1.5 \lambda$, $v_0=\pi/4$, $z_0=0 \lambda$,
$\dot u_0=0.1 \lambda \Gamma$, $\dot v_0=-0.001 \Gamma$, $\dot z_0 =
0.001 \lambda \Gamma$. Notice that the particle dwells on bright
zones of the field.}
\end{figure}
 Using several numerical simulations \cite{Jin}, the correlations
 between the time average value of the atomic
\begin{equation}
\text{l}_{+}\text{l}_{-}= \langle l_{z+}l_{z-}
\rangle_t= m^2\langle[(\vec r +f \hat{e}_x)\times \vec v]_z [(\vec r
-f \hat{e}_x)\times \vec v]_z \rangle_t \nonumber
\end{equation}
 and the $b$ value of light mode were studied.

 As illustrated in Fig.~4, in general,
 the variable $ {l}_{z+}{l}_{z-}$  exhibits
 large fluctuations in an initial transitory stage.
 For paraxial beams, there is a time T such that the time average
 $\text{l}_{+}\text{l}_{-}$ over any interval $t_0 <t<t_0 + T$ becomes
 independent of $t_0$ whenever $t_0>T$. The order of magnitude of
 $T$  is typically $10^4\Gamma$. A rich
structure, consistent with the frequent atomic recoils due to the
light potential wells, could be observed at lower scales. As
expected, the specific numerical results depend on all the involved
variables. Once $q$ and $f$ are fixed, and for the same atomic
initial conditions, a monotonic nonlinear increase of
$\text{l}_{+}\text{l}_{-}$ as a function of $b$ is observed in
general \cite{exception} as illustrated in Fig.~5. For non paraxial
modes the time average $\text{l}_{+}\text{l}_{-}$ is highly
dependent on the time interval considered due to the persistent
increase of the radii of the atom motion illustrated in Fig.~3.

\begin{figure}[ht!]
\includegraphics[width= 0.9 \textwidth]{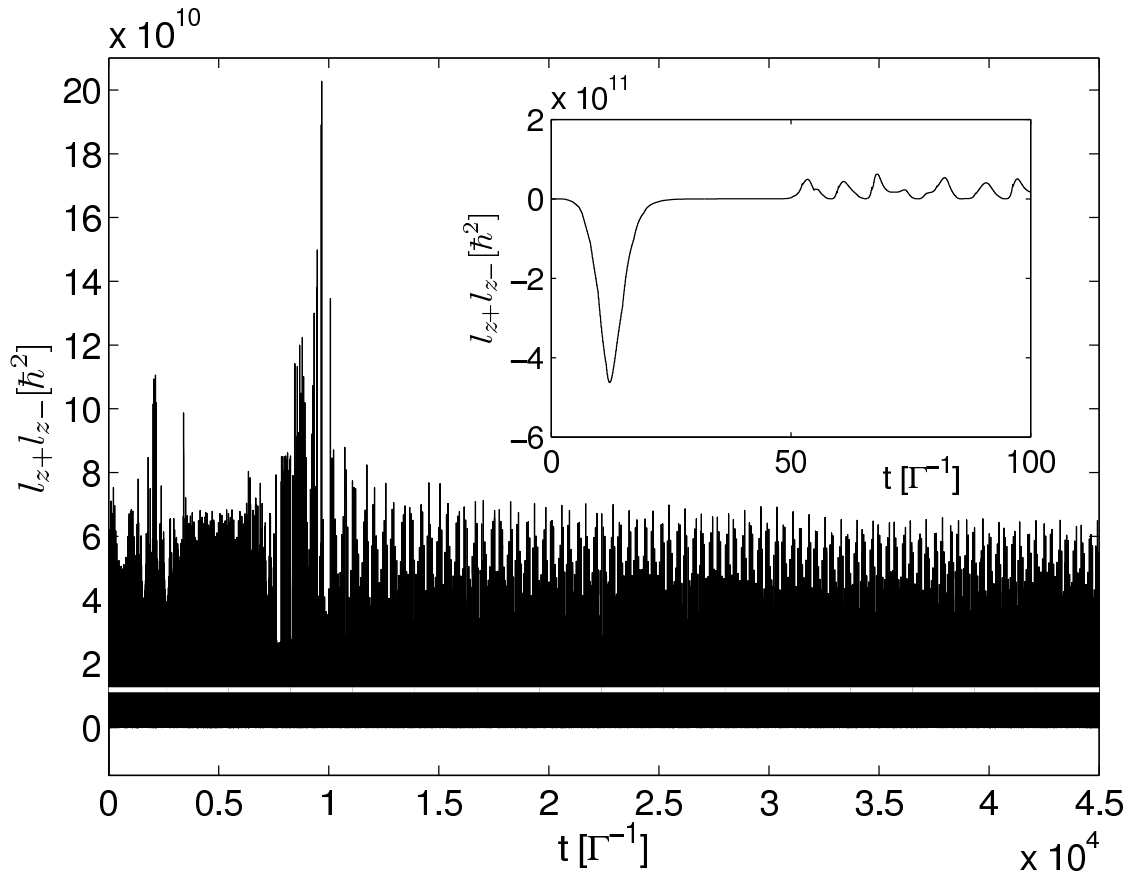}
\caption{(Color Online) Typical behavior of
${l}_{z+}{l}_{z-}$ as a function of time
(black)and its average (white) for paraxial beams. The initial
conditions for the atom are: $u_{0} = 1.2 \lambda$, $v_{0}=\pi/4$,
$z_{0}=0 \lambda$, $\dot{u}_{0}= 0.2 \lambda \Gamma$, $\dot{v}_{0}=
-0.001 \Gamma$ and $\dot{z}_{0} = 0.001 \lambda \Gamma$. The odd
Mathieu beam is of order $n=7$ with half focal distance $f=\lambda$ and
parameter $q =0.1$. The inset graphic is a zoom focused on the
initial behavior of ${l}_{z+}{l}_{z-}$.}
\end{figure}

\begin{figure}[ht!]
\includegraphics[width= 0.9 \textwidth]{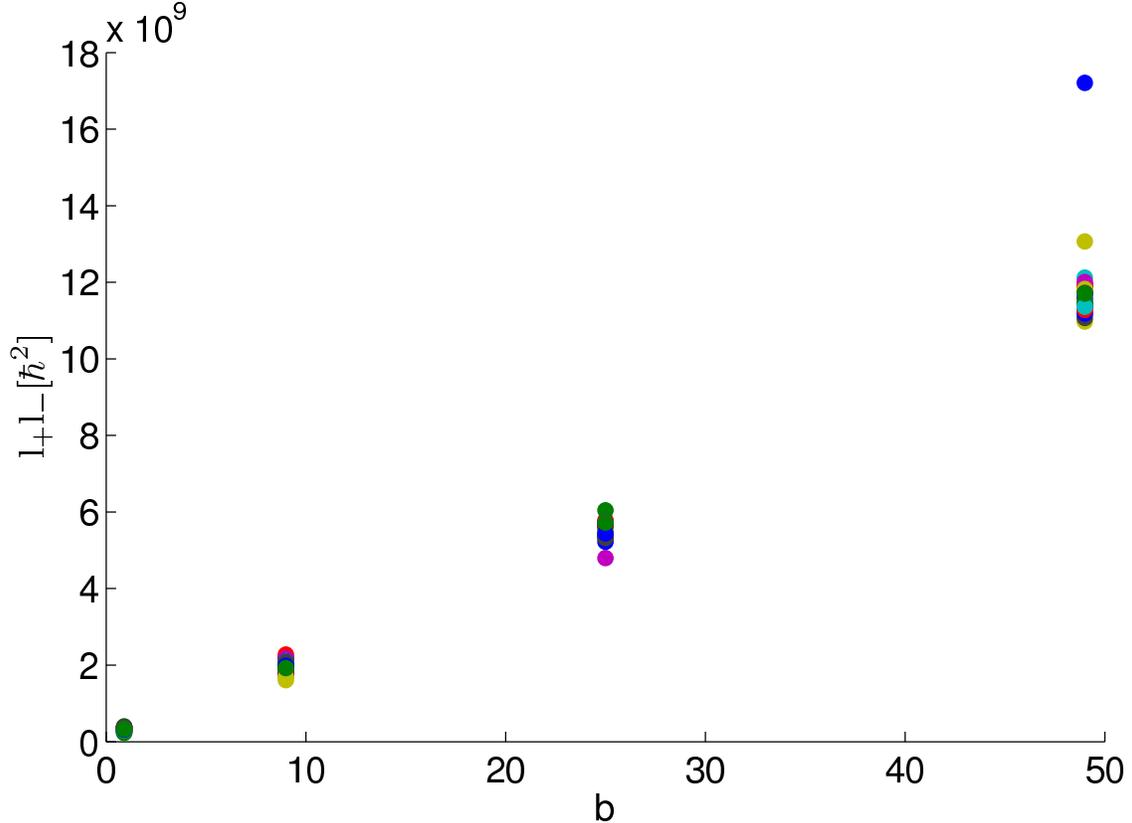}
\caption{(Color Online) Average $\text{l}_{+}\text{l}_{-}$ of the
atom, in units of $\lambda^2\Gamma^{-1}$,  as a function of the $b$
parameter of an odd Mathieu beam of order $n=1,3,5,7$ with half
focal distance $f=\lambda$ and parameter $q =0.1$. The initial conditions
for the atom that yield $\text{l}_{+}\text{l}_{-}$ within the range
plotted here are $u_0 \in [.9,1.5] \lambda$, $v_0 \in
[\pi/16,\pi/4]$, $\dot u_0 \in [.1,.2] \lambda \Gamma$, $\dot v_0 =
-0.001 \Gamma$, $\dot z_0 =0.001 \lambda\Gamma$.}
\end{figure}
\section{Conclusions}

The electromagnetic modes with elliptic-cylindrical symmetry are
characterized by their polarization and the extended set of
parameters $\kappa=\{\omega,k_{z},p,q,b\}$, that is  the field
frequency $\omega$ and wave vector axial component $k_{z}$ related
to the energy $\mathcal{E}$ and $z$ component of the linear momentum
$P_{z}$; the parity $p$ of Mathieu functions and the parameter $q$
which, as in the scalar case, is related to the perpendicular
component of the wave vector in units of the focal distance. It has
been proposed and shown that the transformation generator
$\mathbb{B}$ arising from elliptic symmetry is related to the
dynamical EM variable $\mathcal{B}$ of the electromagnetic field,
giving a physical significance to $b$.

It has been exhibited that the motion of cold atoms in Mathieu beams
can be used to ``measure"  $\mathcal{B}$ since a strong correlation
between the particle product of angular momenta $\text{l}_{+}
\text{l}_{-} $ and the parameter $b$ can be approximately isolated
by using paraxial $TE$ modes in the far off resonance regime.

The use of light beams with elliptical-cylindrical symmetry
has potential applications in controlling the mechanical motion of
atomic systems which could include nanoparticles. Nowadays, it is
well recognized that laser-driven nanoparticles have a variety of
uses in nanofluidics, nanobiotechnology, and biomedicine. However,
although the possibility of optically trapping gold nanoparticles
was demonstrated in 1994 \cite{Svoboda}, developing tweezers for
nanoparticles is not straightforward. The gradient forces with
conventional beams, fall off with particle size. The geometry of
elliptical beams adds the focal distance $2f$ as parameter to tailor
gradient forces besides opening the possibility of using the
mechanical variable $\mathcal{B}$ to select over a wider kind of
motions.

The inclusion and further study of this new dynamical variable could
also help elucidating modern concerns on the ``twisted'' properties
of light, such as why Mathieu functions are intelligent states for
the conjugate pair exponential of the angle--angular momentum
\cite{Hradil2006}.

\begin{acknowledgments}
We thank S. Hacyan and K. Volke-Sep\'ulveda for very useful discussions.
B.M.R.L acknowledges financial support provided by UNAM-DGAPA.

\end{acknowledgments}


\begin{thebibliography}{99}
\bibitem{exception}A remarkable exception corresponds  to the case where
the initial conditions yield an atomic
motion confined in an extremely narrow strip far away from the symmetry axis.
In those cases $\text{l}_{+}\text{l}_{-}$ may
get the same value or even decrease with respect to the average value
 obtained for lower values of the parameter $b$.
Nevertheless this trajectories are very unstable.
For instance, by changing just the initial angular variable $v$ in a $1\%$,
if trapped, the atom acquires a larger   $\text{l}_{+}\text{l}_{-}$ value.
\end{thebibliography}
\end{document}